\documentclass[10pt,a4paper,twocolumn,aps]{revtex4}
\usepackage[utf8]{inputenc}
\usepackage{amsmath}
\usepackage{amsfonts}
\usepackage{amssymb}
\usepackage{graphicx}
\usepackage[dvipsnames]{xcolor}

\begin{document}

\title{
A universal route from avalanches in mean-field models with random fields to stochastic Poisson branching events.
}

\author{Jordi Bar\'o}
\affiliation{
    Departament de Fisica de la Materia Condensada, Universitat de Barcelona, 08028 Barcelona, Spain}
\affiliation{
    Universitat de Barcelona Institute of Complex Systems (UBICS), Universitat de Barcelona, 08028 Barcelona, Spain}
\affiliation{%
Centre de Recerca Matem\`atica,
Edifici C, Campus Bellaterra,
E-08193 Barcelona, Spain
}

\author{\'Alvaro Corral}
\affiliation{Departament de Matem\`atiques,
Facultat de Ci\`encies,
Universitat Aut\`onoma de Barcelona,
E-08193 Barcelona, Spain}
\affiliation{%
Centre de Recerca Matem\`atica,
Edifici C, Campus Bellaterra,
E-08193 Barcelona, Spain
}
\affiliation{Complexity Science Hub Vienna,
Josefst\"adter Stra$\beta$e 39,
1080 Vienna,
Austria
}

\begin{abstract}

Avalanches in mean-field models can be mapped to memoryless branching processes defining a universality class. 
We present a reduced expression mapping a broad family of critical and subcriticial avalanches in mean-field models at the thermodynamic limit to rooted trees in a memoryless Poisson branching processes with random occurrence times. 
We derive the exact mapping for the athermal random field Ising model and the democratic fiber bundle model, where avalanche statistics progress towards criticality, and as an approximation for the self-organized criticality in slip mean-field theory. Avalanche dynamics and statistics in the three models differ only on the evolution of the field density, interaction strength, and the product of both terms determining the branching number. 

\end{abstract}


\maketitle

Many complex systems with discrete symmetry breaking exhibit \textit{avalanche dynamics}.
Under quasistatically slow external driving, the temporal evolution $v(t)$ of physical observables appears to be split between long periods of quiescence and well-delimited fast transformation events, the so-called \textit{avalanches}. Due to their fast nature, such avalanches can be regarded as instantaneous and labeled as point events $k$ in time with a defined time origin $t_k$, and marked by their magnitude or size $
{S_k = \int_{t_{k}}^{t_k+T_k}v(t)dt}
$, and by the duration $T_k$ of the excursion out of quiescence, which models the dynamics of the system. 
Many physical systems exhibit criticality in the form of scale-free avalanches \cite{Bak1987,Bak1991,Sornette1992,Garcia-Pelayo1994} rendering power-law distributions of sizes and durations. 

The cascading nature of avalanches has long been associated with branching processes \cite{Alstrom1988,Zapperi1995,Grassberger2021,Christensen1993, Garcia-Pelayo1994, Corral2018,LeDoussal2022} where each element $i$ in a generation $g$ triggers a number $Y_{i,g}=0,1,2..$ of new elements forming part of generation $g+1$. The total number of offspring in each generation $g$ as given by $Z_g=\sum_{i=1}^{Z_{g-1}} Y_{i,g-1}$, imposing $Z_0=1$.
Avalanche sizes ($S$) are analogous to the sizes of extinguished rooted trees $\Delta:=\sum_{g=0}^{G-1} Z_g$, after a number of generations, or \textit{depth}, $G$, analogous to durations ($T$), such that $Z_{G-1}> 0$ and $Z_G=0$.
 We focus on those processes, customarily designated to belong to the mean field (MF) universality class (MF-UC), where avalanche propagation can be mapped to memoryless, a.k.a. Galton-Watson, branching processes~\cite{Alstrom1988,Zapperi1995,Grassberger2021} where all $Y_{i,g}$ are i.i.d. (independent and identically distributed).
 We highlight the properties of memoryless branching constituting the MF-UC:\\
(i) The expected number of elements in generation $g$ reads
$\mathrm{E}(Z_{g}) = \mathrm{E}(Y)^{g}=\mu^{g}$,
where $\mu :=\mathrm{E}(Y)$
is the branching parameter.
Then, the expected tree size is:
\begin{equation}
\mathrm{E}
( \Delta ) 
 = \sum_{g=0}^{\infty} \mathrm{E}(Z_g)
= \sum_{g=0}^{\infty} \mu^g = \frac{1}{1-\mu},
\label{eq:Deltamu}    
\end{equation}
for $\mu \le 1$, whereas $E(\Delta) \rightarrow \infty$ for $\mu\ge 1$.\\
(ii) For any offspring distribution with well-defined statistical moments and
at criticality (that is, for $\mu=1$),
the survival probability beyond generational depths $G$ is asymptotically
$P(G_k> G) \propto G^{-1}$, rendering a probability mass function (p.m.f.) 
 of avalanche depths \cite{Harris_original}: 
\begin{equation}
\label{eq:pT}
 \mathit{pmf}_{G}(G) \sim G^{-\kappa_G} \;\;\mathrm{with} \;\; \kappa_G=2.
\end{equation}
 (iii) Close to criticality \cite{Aldous1991b,Pitman2006}, the size of such extinguished rooted trees ($\Delta$), is linked to characteristic avalanche depths, e.g., $G$, in a relation of the type: 
\begin{equation}
\label{eq:TDelta}
\mathrm{E}( G| \Delta  ) \sim \Delta^{\gamma}  \;\;\mathrm{with} \;\; \gamma = 0.5,
\end{equation}
leading to the critical size distribution:
\begin{equation}
\label{eq:pDelta}
 \mathit{pmf}_{\Delta}(\Delta) \sim \Delta^{-\kappa_\Delta}  \;\;\mathrm{with} \;\; \kappa_\Delta = 1+ \gamma (\kappa_G - 1) = 3/2.
\end{equation}

These memoryless branching (from now on MF) exponents, are ubiquitous in natural avalanche process close to criticality \cite{Beggs2003, Dahmen2017a, Denisov2017, Baro2018b} and are reproduced by MF models \citep{Bak1987,
Alstrom1988,Zapperi1995,Manna1991,Grassberger2021,Hemmer1992,Sethna1993,Dahmen2009}.
As a paradigmatic example, sand-pile models with random grain redistribution rules leading to self-organized criticality (SOC) have traditionally been modeled as binomial branching processes \cite{Bak1987,Manna1991,Grassberger2021}.
In this Letter we identify another category of avalanche models which maps to a Poisson branching process.
Inspired by early results by Sethna et al. \cite{Sethna1993} we derive a universal route to Poisson branching processes from (i) the mean-field \textit{random field Ising model} (RFIM) and prove its validity for two other prototypical models: (ii) \textit{the democratic fiber bundle model} (DFBM)~\cite{Hemmer1992, Kloster1997}, and (iii) \textit{slip mean-field theory} (SMFT)~\cite{Dahmen2009}.\\

\begin{figure*}
\includegraphics[width=2\columnwidth]{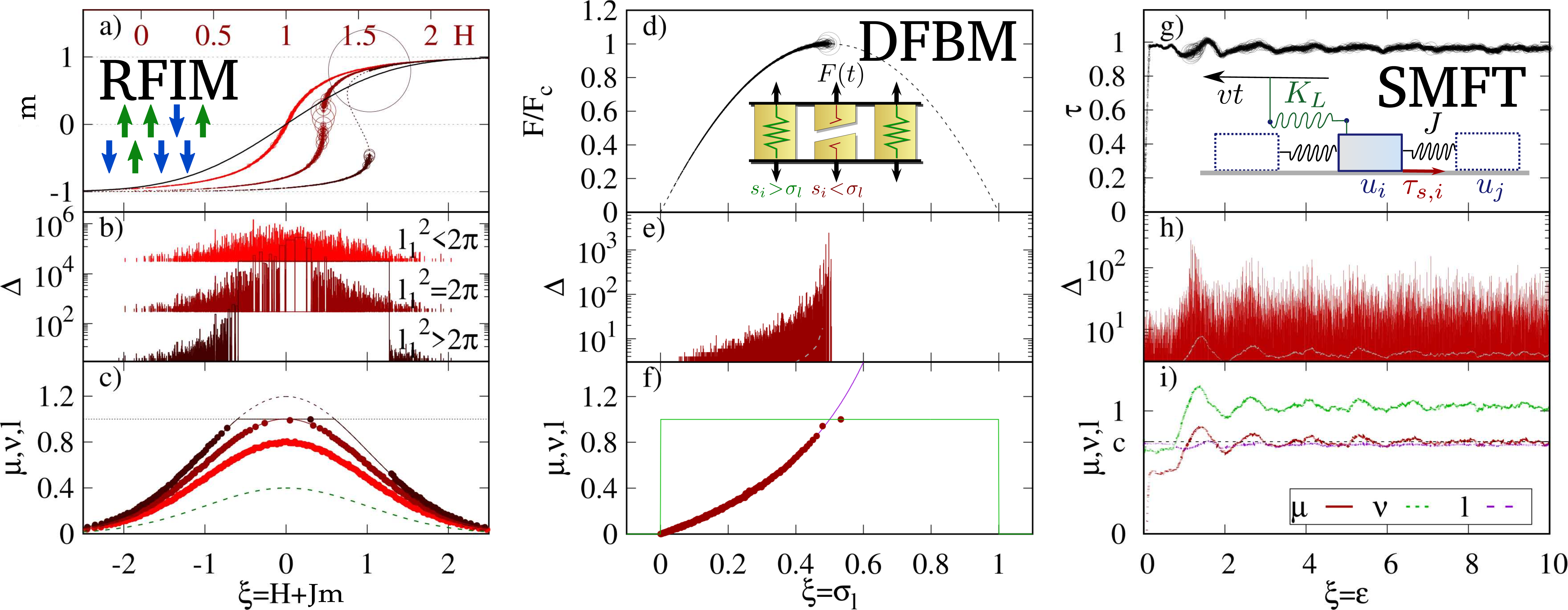} 
\caption{\label{fig:models} Numerical simulations and evolution of parameters in terms of the mean field $\xi$: (a-c) Three RFIM with $N=5\times 10^4$ spins and subcritical ($J=l_1/2 = 1$), critical ($J=\sqrt{\frac{\pi}{2}}$) and supercritical ($J=1.5$) Gaussian disorder field; 
(d-f) a DFBM of $n(0)=5\times 10^5$ with uniform distribution of strengths; and  (g-i) a SMFT with $N= 10^5$ and a broad uniform distribution of disorder ($\tau_{s,i}\in (0.675, 1.325)$) and $c=0.75$. Top row shows the evolution of the traditional order parameters. An additional x-axis has been added to (a) to show the dependence in terms of external field $H$ for different values of disorder (red lines colorcoded as (b)). Middle row shows the sequence of avalanche sizes ($\Delta>3$). Bottom rows show the selected (c,f) and emergent (i) density $\nu$ of the thresholds, $l_1$ (constant and eluded in c) and the estimated $\hat{\mu}$ (\ref{eq:muML}) using 1000 avalanches. 
}
\end{figure*}

\textit{i) Tunable branching in the random field Ising model:}
Let us write the Hamiltonian of a system of 
 $N$ spins with value $s_i=\lbrace -1, +1 \rbrace$,  and all-to-all ferromagnetic (${J>0}$)  interactions in reduced units as: 
\begin{equation}
\mathcal{H}(t)=-\sum_{i=1}^N s_i (\xi(t)-h_i),
\label{eq:RFIMhamiltonian}
\end{equation}
where the quenched disordered fields $h_i$ uniquely felt by each spin $i$
are i.i.d.
given a probability density $\rho_h(h_i=\xi')d\xi'$.
The global term, or \textit{mean field}, $\xi(t)$ is the same for all spins and
is given by a linear combination of the external driving field $H(t)$ and the interaction with all other spins $s_j$ approximated as $ \frac{J}{N} \sum_{j\neq i}s_j\approx  \frac{J}{N} \sum_{j}^{N}s_j =:J m $, a.k.a. the magnetization field, 
such that: 
\begin{equation}
\xi(t) := H(t) + J m(t).
\label{eq:RFIMfield}
\end{equation}
Under athermal conditions, $\mathcal{H}(t)$ minimizes locally by setting each spin as 
$s_i= sign\;[  \xi(t) - h_i ]$.
Considering a monotonic driving $\dot{H}>0$ from an original configuration $\lbrace s_i \rbrace = \lbrace -1 \rbrace$, each spin flips on the condition: $h_i < \xi(t)$.
Therefore, $m=m(\xi)$ since it is proportional to the number of spins with $h_i>\xi$. This intrinsic relation potentially leads to an indeterminate $m$ above a critical density $ \rho_h(\xi)>(2J)^{-1}:=\rho_c(J)$~\citep{Sethna1993} (see Fig.~\ref{fig:models}.a). The stochastic  $\lbrace h_i \rbrace$ introduces shorter indeterminate intervals below $\rho_c$ (Fig.~\ref{fig:models}.b), corresponding to avalanches.\\

Inspired by the study of avalanches in the RFIM and random transport ~\citep{Sethna1993,Fisher1998}, we construct here a general model consisting of a sorted set of random thresholds $\lbrace .. X_i < X_{i+1} .. \rbrace$, e.g. random fields $h_i$ in the RFIM, and a mean field $\xi(t)$ defining the state of the system as:
\begin{equation}
\label{eq:RTM}
\xi(t) := \mathcal{B}(t) + \sum_{i|X_i<\xi(t)} \mathcal{R}_i(t-t_i),
\end{equation}
that is, a linear combination of a monotonic  external driving term $\mathcal{B}(t)$
and monotonic responses $\mathcal{R}_i(\tau)$ caused by each activated threshold. 
The activation time of each threshold ($t_i$) is the instant when the increasing field
$\xi(t)$ crosses the threshold $X_i$
assuming
$\mathcal{R}_i(\tau<0)=0$.\\

The nonlinear coupling between $\xi$ and $\mathcal{R}_i(\tau)$ gives rise to avalanches. If the driving is quasistatically slow compared to the response $\dot {\mathcal{B}} \ll  \dot{\mathcal{R}_i}$, avalanches are well defined in time as concatenated responses $\lbrace \mathcal{R}_i \rbrace$ which we can consider to be instantaneous, such that:
\begin{equation}
\mathcal{R}_i(\tau) = l_i \Theta(\tau-\delta_g),
\label{eq:R0}
\end{equation} 
after an infinitesimal delay time $\delta_g$ clustering field steps in discretized generations, a.k.a. shells, $g$ within the avalanche.
Here, $\Theta(x)$ is the step function defined as $\Theta (x)= 0$ for $x \leq 0$ and $\Theta (x)= 1$ for $x > 0$.
The factor $l_i$ is a characteristic field increase in units of $\xi$. The stochastic nature of avalanches derives from the i.i.d. $\left\lbrace{ X}\right\rbrace$, with an expected number $\nu$, a.k.a. density, of thresholds per unit of field ($\xi$). The match between the RFIM and the model defined by Eqs. (\ref{eq:RTM}) and (\ref{eq:R0}) is directly given by:
 \begin{equation}
l_i  \equiv \frac{2 J}{N };\;
\dot{\mathcal{B}}(t) \equiv  \dot{H}(t);\;
\lbrace X_i \rbrace \equiv  \lbrace h_i \rbrace,
\label{eq:RFIM}
\end{equation}

\begin{figure}
\includegraphics[width=\columnwidth]{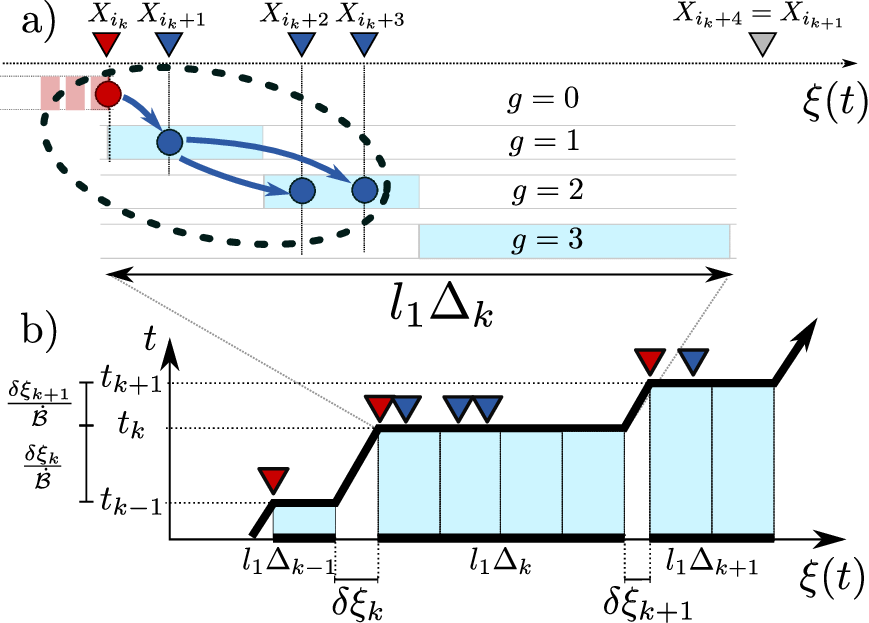} 
\caption{\label{fig:RTM_A} (a) One avalanche represented as a branching process within (b) an avalanche sequence represented as a point process in field $\xi$ and time $t$ considering constant driving  rate $\dot{\mathcal{B}}$.
Red triangles represent the values of thresholds ($i_j$) triggering each avalanche $j$ when $\xi(t_j)=X_{i_j}$. Blue triangles represent thresholds activated within the avalanche propagation. Blue rectangles  of length $Z_g l_1$ represent the interval of instant transformation at each shell $g+1$ leading to the activation of $Z_{g+1}$ thresholds $\lbrace{X_i}\rbrace$. Blue arrows represent causal connection, i.e., branching, between activations (circles). 
}
\end{figure}

Finite avalanches themselves are too small to modify the state $\xi(t)$ of infinite systems.  By considering that $\nu(\xi)$ and $l_i(\xi)$ are, at most, function of state,  these can be approximated as constants during intervals of avalanche sequences.
Such infinitesimal intervals of $\xi$ variation can be mapped onto a stochastic point process with stationary laws, as illustrated in Fig.~\ref{fig:RTM_A}.b.
Avalanches are well identified sudden advancements of $\xi(t)$, separated by i.i.d. time intervals of width $\delta \xi_k/\dot{\mathcal{B}}$,
with $\delta \xi_k$ the change in $\xi(t)$ between the end of the $k-1$-th avalanche and the beginning of the $k$-th. The size and duration marking the event is determined by the propagation profile between the nucleation and halting of an avalanche, 
as illustrated for avalanche $k$ in Fig.~\ref{fig:RTM_A}.a.
The avalanche is triggered through quasistatic driving $\dot{\mathcal{B}}(t) > 0$ when $\xi(t)$ reaches the threshold value of the element $X_{k_0}$ nucleating the avalanche $k$ at an instant $t_{k,0}$. 
Since we consider the immediate response approximation (\ref{eq:R0}), the external field $\mathcal{B}(t)$ will remain constant during the propagation of the avalanche. 
The variation of $\xi$ from $t_{k,0}$ to $t_{k,1}= t_{k,0}+\delta$ reads $\xi(t_{1}) = \xi(t_{0}) + l_1$, crossing a number $Z_{k,1}$ of consecutive thresholds $\lbrace X_{k_0+1}, .., X_{k_0+Z_{k,1}}\rbrace$, which will constitute the shell, or generation, $g=1$ of event $k$. 
Since $X_i$ are i.i.d. and homogeneous, the number of ``offspring thresholds'' activated within the interval $\Delta \xi = l_1$ is a Poisson number with parameter:
\begin{equation}
\mu := E(Z_{k,1}) \equiv E(Y)=\nu \times l_{1}.
\label{eq:branchingMF}
\end{equation}

 In its turn, the activation of the $Z_{k,1}$ thresholds will advance $\xi(t)$ over the next instant: $\xi(t_{k,2}=t_{k,0}+2\delta ) = \xi(t_{k,1}) + Z_{k,1} l_{1}$ which will cross a Poisson number $Z_{k,2}$, with $E(Z_{k,2})= Z_{k,1} \mu$, of thresholds $\lbrace  X_{k_0+Z_{k,1}+1}, .., X_{k_0+Z_{k,1}+Z_{k,2}} \rbrace $ that comprise the shell $g=2$. 
 The process stops when $Z_{k,G}=0$, resulting in an avalanche of size $\Delta_k := 1+\sum_{g} Z_{k,g}$ with a corresponding increase in field $\xi(t_{k,G}) = \xi(t_{k,0}) + \Delta_k l_1$ and lasting a time $T_k=G_k \delta_g$.
 As described, the propagation of the avalanche is equivalent to a memoryless Poisson branching process with a branching ratio $\mu$. 
Subcritical ($\mu<1$) avalanche sizes ($\Delta$) are equivalent to subcritical Poisson tree sizes, which are i.i.d. according to a Borel distribution \cite{Otter1949, Pitman2006}:
\begin{figure}
    \centering
    \includegraphics[width=\linewidth]{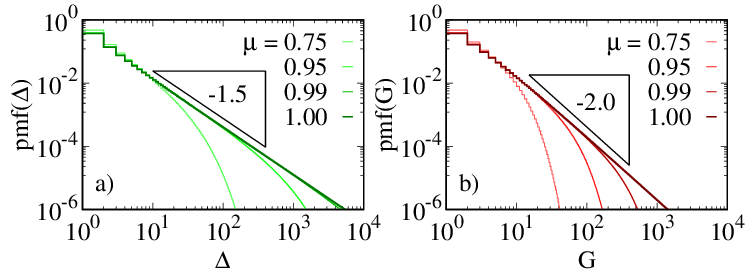}
    \caption{Exact local distributions of (a) sizes and (b) durations given by (\ref{eq:Borel}) and (\ref{eq:Tdist}) valid for any model represented by (\ref{eq:RTM}).}
    \label{fig:distros}
\end{figure}

\begin{equation}
 \mathit{pmf}_{\Delta}(\Delta;\mu) = \frac{(\mu \Delta)^{\Delta-1} e^{-\mu \Delta}}{\Delta !},
\label{eq:Borel}
\end{equation}
with $\Delta=1, 2...$;
Moreover, avalanche survival after $G$ generations
is given by the self-recurrent expression~\cite{Otter1949}:
\begin{equation}
 P_G(g;\mu) = 1- e^{-\mu P_G(g-1;\mu)} {\;\;;\;\;} P_G(0;\mu) = 1.
\label{eq:Tdist}
\end{equation}

As shown in Fig.~\ref{fig:distros}, for large $\Delta$ and $G$ and close to $\mu\sim 1$, these expressions can be numerically approximated by the universal forms (\ref{eq:pT}) and (\ref{eq:pDelta}).
The maximum likelihood estimation for $\mu$ from (\ref{eq:Borel}) corresponds to (\ref{eq:Deltamu}):
\begin{equation}
\hat{\mu} = 1- \langle \Delta \rangle^{-1},  
\label{eq:muML}
\end{equation}
with $\langle \Delta \rangle$ the sample mean of the avalanche size,
and with the uncertainty given by the Fisher information:
$\widehat{\sigma^2_\mu} = {\mu(1-\mu)}/{{N}}.  $\\

We remark that the dynamics and statistics of avalanches are fully determined by only two terms: (i) the density of random  thresholds ($\nu$) determining the rate of avalanches in units of $\xi$; (ii) the step size ($l_i$) converting branching numbers into units of $\xi$. 
Given constant $l_1$ and $\nu$, the resulting avalanche sequence can be exactly modeled as an homogeneous point process in the $\xi$ axis ($x-$axis in Fig.~\ref{fig:RTM_A}.b) with dead-times $l_1 \Delta_k$ being the sizes $\Delta_k$ i.i.d. Borel distributed with parameter $\mu$ (\ref{eq:branchingMF}). 
Considering a slow driving rate $\dot{\mathcal{B}}$, the dead times are negligible, recovering a standard Poisson process in time ($y-$axis in Fig.~\ref{fig:RTM_A}.b).

The representation in (\ref{eq:RTM}) is valid in general for non-stationary $\lbrace l_1, \nu \rbrace (\xi)$ at the thermodynamic limit, potentially leading to a divergence of susceptibilities at $\mu=1$ with a corresponding third independent exponent. E.g., Fig.~\ref{fig:models}.a-c, show transitions between different avalanche regimes in the RFIM given the same non-uniform $\nu(\xi)$ from a Gaussian distribution, and different values of $J$. Sizes diverge at criticality ($\mu=1$) matching the global instability limit at $\rho_c(J)$, before entering supercritical regimes which are out of the scope of this Letter.\\

\textit{ii) Critical failure in fiber bundle models:}
The \textit{democratic fiber bundle model} (DFBM) is a prototype model of so-called critical failure~\cite{Hemmer1992,Kloster1997}, meaning that, as the driving progresses, avalanche statistics exhibit a monotonic evolution towards and beyond criticality, $\mu=1$ in terms of (\ref{eq:RTM}), causing a catastrophic failure or instability.
The DFBM simulates the deformation and resilience of an ensemble of brittle elastic elements, a.k.a. fibers, plugged in parallel and sustaining an increasing load. Each fiber fails under stress driving when the local stress $\sigma_l$ exceeds a random strength $s_i$, which is i.i.d. given a density $\nu(\sigma_l):=\frac{dn}{d\sigma_l}$. The total tension ($F(t)$) is equally sustained by the number of unbroken fibers $n(t)$. 
Therefore, the local stress for each fiber reads $\sigma_l(t)=F(t)/n(t)$. Assuming a monotonous driving in $F(t)$, $n(t)$ is the number of fibers $\lbrace i \rbrace$ with $s_i>\sigma_l(t)$. It is then convenient to redefine $n(\sigma_l) := n(t| s_i>\sigma_l(t)) = \int_{\sigma_l(t)}^{\infty} \nu(s')ds'$. The mechanical stability of the bundle is defined by the constitutive equation:
\begin{equation}
\sigma_l = \frac{F}{n(\sigma_l)}.
\label{eq:DFBM}
\end{equation}
 Under force driving, (\ref{eq:DFBM}) displays a singular instability  at the failure point $\sigma_{l,f} = n(\sigma_{l,f})/\nu(\sigma_{l,f})$, or $F_f = n(\sigma_{l,f}) \sigma_{l,f}$, above which the bundle fails catastrophically (see Fig.~\ref{fig:models}.d). Defining the mean field as $\xi(t) \equiv  \sigma_l(t)$, and considering large system sizes such that $n\approx n-1$, (\ref{eq:DFBM}) can be rewritten as the sum of the driving intervals and the steps $\sum_i [\sigma_l(t_i+\delta t)-\sigma_l(t_i)$] caused by fibers breaking on the thresholding $\sigma_l>s_i$, leading to a form equivalent to (\ref{eq:RTM}) with terms:
\begin{equation}
l_i  \equiv \; \frac{F(t_i)}{[n(t_i)]^2};\;
\dot{\mathcal{B}}(t)  \equiv  \; \frac{\dot{F}(t)}{n(t)};\;
\lbrace X_i \rbrace  \equiv \; \lbrace s_i \rbrace.
\label{eq:mfDFBM}
\end{equation}
In the DFBM, $l_i$ and $\dot{\mathcal{B}}$ are non-stationary and depend on the state of the fracture process. In particular, we observe a mild increase in the activity over time, and the failure point $(\sigma_{l,f},F_f)$ coincides with the critical branching ratio $\mu_f =  \sigma_{l,f} \nu(\sigma_{l,f})/n(\sigma_{l,f})= 1$ 
(Fig.~\ref{fig:models}.f), reached after a monotonic increase in $\mu$. This equivalence between failure and avalanche criticality exposed through (\ref{eq:RTM}) vindicates the use of raw avalanche statistics, e.g., acoustic monitoring, to predict failure \cite{Main1990,Lockner1993,Kun2014,Baro2018b,Diksha2022,Diksha2024}. Below $F_f$, fibers fail in subcritical bursts, following a Poisson branching process with Borel i.i.d. sizes (\ref{eq:Borel}) for $\mu(\sigma_l)<1$, matching the solution of the so-called forward condition introduced by C. Hemmer and A. Hansen~\cite{Hemmer1992}.\\

\textit{iii) Self-organized criticality in frictional sliding:}
The model representation in (\ref{eq:RTM}) is still adequate to describe systems where $\lbrace l_1, \nu \rbrace$ cannot be exactly reduced to constants nor even function of state.
We introduce the so-called \textit{slip mean field theory} (SMFT)~\cite{Fisher1996, Dahmen1998, Dahmen2009} as a prototype for the emergence of SOC in dissipative systems with dynamic disorder. The SMFT provides a theoretical description of frictional sliding and non-afine plastic deformation restricted to positive stress transfer, with a scale of applicability ranging from fault systems to dislocations in bulk metallic glasses \cite{Ben-Zion2011,Dahmen2011, Friedman2012, Dahmen2017a, Denisov2017}. The SMFT models representative elements of a rough surface as a system of elastically coupled frictional solid blocks  \cite{Dahmen2009}. At time $t$ each block is arrested at a displacement position $u_i(t)$ due to a static frictional force $\tau_{s,i}$, and interacts elastically with a common driving element displacing at speed $v$, with elastic coefficient $K_L$, and a set of neighboring sites with displacements $u_j(t)$, with an elastic coefficient $J$ (see sketch in Fig.~\ref{fig:models}.g).
Element $i$ slips when the shear stress
$\tau_i(t) = J/k \sum_{\langle ji \rangle}^k (u_j(t)-u_i(t)) + K_L (v t - u_i(t))$ surpasses the arresting strength $\tau_{s,i}$.
By considering $k=N$, i.e, the system size, the MF interaction term $ J/N \sum_{\lbrace j\rbrace} (u_j- u_i) = J(\overline{u}-u_i)$. The mechanics in MF approximation reads:
\begin{equation}
\tau_i = J\; \overline{u}\;+ K_l v t - (J/N+K_L)u_i,
\label{eq:SMFTbase}
\end{equation}
The stress released $\tau_i \to \tau_i - \delta \tau_i$ is partially transferred to the rest of the elements $j \neq i$ as: $\tau_j \to \tau_j + c/N \delta \tau_i$, with a fraction determined by the conservation term $c= \frac{J}{J+K_L}$. By isolating all stochastic terms according to the new position ($u_i$) in a single term $X_i$, we can express the dynamics as (\ref{eq:RTM}) by considering: 
\begin{equation}
l_i  \equiv  \frac{c\delta \tau_i}{N}  ;\;
\dot{\mathcal{B}}(t)  \equiv   K_L v;\;
\lbrace X_i \rbrace  \equiv \lbrace \tau_{s,i}+ (K_L+J/N)u_i\rbrace. 
\label{eq:SMFT}
\end{equation}
Here, $l_i$ can differ by construction for each $i$ and the threshold values $\lbrace X_i \rbrace$ are deterministically resampled after each  slip in $i$ and self-organized towards a regular distribution, approximately uniform $U(0,\overline{\tau_{s,i}})$ for low disorder. 
The effective $\mu$ is therefore site dependent and the resulting branching process is not strictly Poisson, yet close to it for low disorder in $\lbrace \tau_{s,i} \rbrace$. Fig.~\ref{fig:models}.g-i, shows the evolution of the moving averages ($\overline{l_i}(\xi),\overline{\nu}(\xi)$)  in numerical simulations. 
After a transient regime, the estimated parameters fluctuate around $\overline {l_i} \approx c \langle \tau_s \rangle /N$, since $\delta\tau_i \approx \tau_{s,i}$, and $\overline{\nu} \approx  N \langle \tau_c \rangle^{-1}$, leading to the self-organized branching parameter $\langle \mu \rangle = c$ which is critical at the conservative limit $c=1$.\\

We believe that the coincidental representation of type (\ref{eq:RTM}) for the three selected models extends to other constructions defined in quasistatic and athermal conditions, where avalanches emerge from discrete instabilities in a continuous mean field $\xi$ driven across a discrete field of i.i.d. random thresholds. This specific route towards a Poisson branching is exact at subcritical regimes. Under conditions $l_1=l_1(\xi)$ and $\nu=\nu(\xi)$ subcritical avalanche dynamics and statistics are non-stationary, with instant size and duration probabilities matching exactly (\ref{eq:Borel}) and (\ref{eq:Tdist}), albeit with state-dependent $\mu(\xi)$, and (\ref{eq:muML}) provides a model-free distance to criticality. 
Excluded from the MF-UC altogether are those mean-field models with scale-free threshold distributions, such as pseudo-gaps in elastoplastic models~\cite{Hebraud1998,Karmakar2010,Nicolas2018}, since the $\lbrace X_i \rbrace$ values cannot be considered asymptotically homogeneous. Excluded are also those systems with short-range interactions or any type of depth-dependence, i.e., memory, in the branching. We finally recall that the concept of MF-UC extends beyond the condition of all-to-all interactions. Such systems commonly underlay branching processes without, or little, memory. E.g., randomized remote interactions, as in the aforementioned sand-pile models, the embedding in structural small-world networks \citep{Beggs2003,Denisov2016,Baro2021}, the self-organized pruning of loops in directed networks \citep{Arcangelis2018}, or other mechanisms breaking interaction loops validate (\ref{eq:Deltamu}),(\ref{eq:pT}),(\ref{eq:TDelta}) and (\ref{eq:pDelta}) with MF exponents, but different offspring distributions.

\acknowledgements{ 
J.B. acknowledges financial support from project PID2022- 136762NA-I00 funded by MICIU/AEI/10.13039/501100011033 and FEDER, UE.
A.C. acknowledges PID2021-125979OB-I00
as well as the Severo Ochoa and Mar\'{\i}a de Maeztu Program for Centers and 
Units of Excellence in R\&D of the Spanish AEI (CEX2020-001084-M).
}
\bibliography{biblioALL}

\end{document}